# $p$ Orbital flat band and Dirac cone in the electronic honeycomb lattice

*Thomas S. Gardenier[1‡], Jette J. van den Broeke[2‡], Jesper R. Moes[1], Ingmar Swart[1], Christophe Delerue[3], Marlou R. Slot[1†], Cristiane Morais Smith[2], Daniel Vanmaekelbergh[1\*]*

[1]Debye Institute for Nanomaterials Science, Utrecht University, The Netherlands.

[2]Institute for Theoretical Physics, Utrecht University, The Netherlands.

[3]Université de Lille, CNRS, Centrale Lille, Yncréa-ISEN, Université Polytechnique Hauts-de-France, UMR 8520 - IEMN, F-59000 Lille, France

Theory anticipates that the in-plane $p_x$, $p_y$ orbitals in a honeycomb lattice lead to new and potentially useful quantum electronic phases. So far, $p$ orbital bands were only realized for cold atoms in optical lattices and for light and exciton-polaritons in photonic crystals. For electrons, in-plane $p$ orbital physics is difficult to access since natural electronic honeycomb lattices, such as graphene and silicene, show strong $s - p$ hybridization. Here, we report on electronic honeycomb lattices prepared on a Cu(111) surface in a scanning tunneling microscope that, by design, show (nearly) pure orbital bands, including the $p$ orbital flat band and Dirac cone.

The electronic properties of two-dimensional solids, including materials with Dirac bands and topological insulators, are largely determined by the geometry of the atomic lattice and the nature of the interacting orbitals. [1, 2] A compelling case is presented by the system of in-plane $p_x$, $p_y$ orbitals in a honeycomb lattice providing an electronic flat band, due to geometric frustration, and a $p$ type Dirac cone. [3-5] The in-plane $p$ orbitals in the trigonal honeycomb lattice cannot form conventional bonding – antibonding combinations; their interaction gives rise to complex interference patterns. As a result, the four in-plane $p$ bands consist of a non-dispersive flat band, followed by two dispersive bands forming a Dirac cone at higher energy, followed by another flat band. Intrinsic spin-orbit coupling will open a gap at the Dirac point (the quantum spin Hall effect) and detach the flat band from the Dirac cone, making it topological. [6, 7] Since the kinetic energy is quenched in the flat band, the dominant energy scale is set by interactions. It has been predicted that this will lead to new quantum phases, such as unconventional superconductivity and Wigner crystals. [4, 8] The physics of in-plane $p$ orbitals has been studied with ultracold atoms in optical lattices, [8-12] light in photonic systems, [13] and exciton-polaritons in a semiconductor pillar array. [14, 15]. However, an experimental realization of an electronic material in which the physics of in-plane $p$ orbitals can emerge by design has not yet been reported.

Natural electronic honeycomb systems show interesting results, but there is considerable hybridization between different types of orbitals. [6] In graphene, the most studied electronic honeycomb lattice, the $s$- and in-plane $p_x$, $p_y$ orbitals of the carbon atoms hybridize and form $sp_2$ electronic bands, the lower one being completely filled. [3] This filled band leads to a very strong in-plane bonding between the carbon atoms, but is not electronically active. The remaining $p_z$ orbitals (perpendicular to the graphene plane) form $\pi$ bonds, resulting in two bands touching at the (K, K') Dirac points at which the Fermi energy is situated. The linear energy-wave vector dispersion (Dirac cone) around the (K, K') points is responsible for the exciting electronic properties of graphene. [3]

Here, we report solid-state designs for electrons in which the physics of in-plane $p$ orbitals fully emerge. Our work is inspired by the first reported artificial electronic honeycomb lattice, [16] based on the surface state electrons of a Cu(111) surface. Extending this concept, we design honeycomb lattices consisting of atomic sites with a variable degree of quantum confinement, and electronic coupling between them. Muffin-tin calculations show that it is possible to create lattices in which the on-site $s$ orbitals and $p$ orbitals are sufficiently separated such that Dirac-cones and a flat band emerge with nearly pure orbital character. The band structure is experimentally investigated by measurement of the local density of states and wave function mapping.

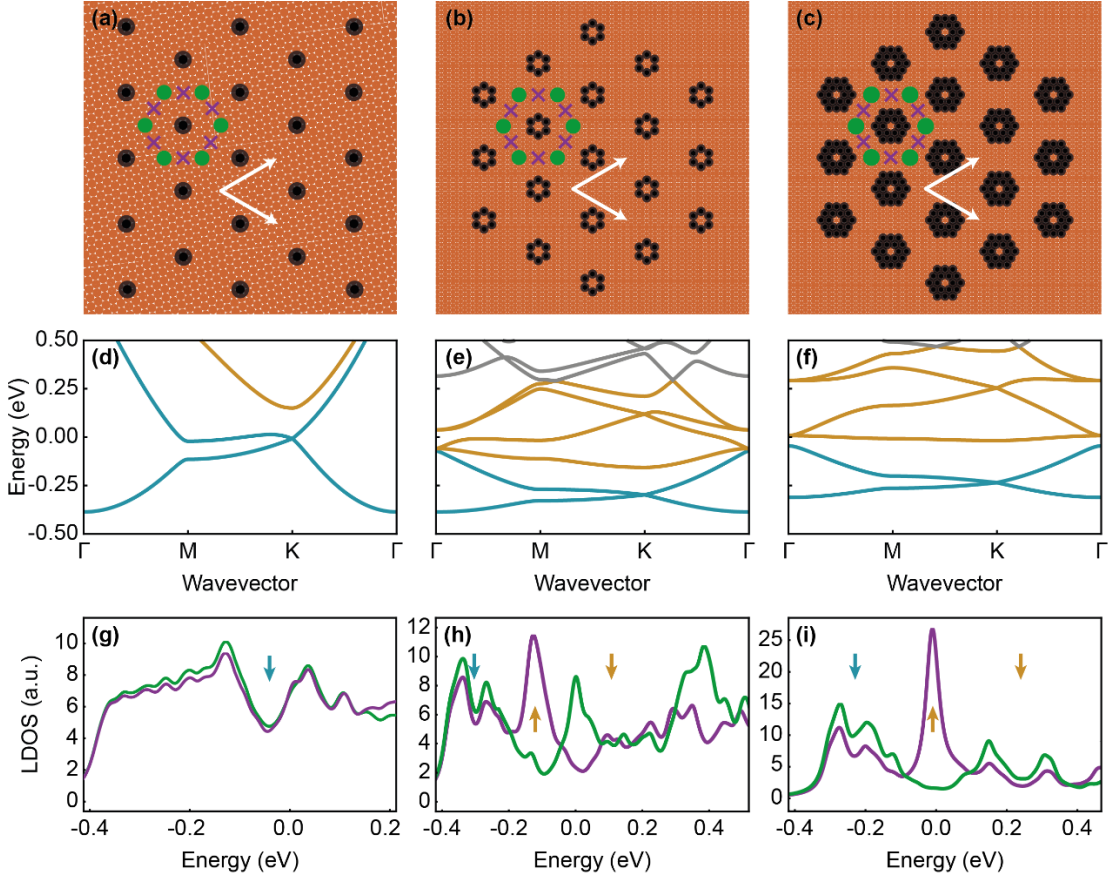

**Figure 1.** Designs for artificial atoms in a honeycomb lattice, corresponding band structures and LDOS spectra.

*(a-c) Schemes of a Cu(111) surface (copper) and the positions of the CO molecules (black) defining the on-site energies of the s- and p orbitals and their inter-site coupling. The lattice sites are indicated in green, the bridge sites with purple crosses. (a) The lattice reported by Gomes et al,[16] with a honeycomb lattice vector of 1.92 nm. (b) Lattice with single-ringed CO rosettes as scattering islands and a honeycomb lattice vector of 3.58 nm, corresponding to 14 Cu atoms, (c) lattice with double-ringed CO rosettes as scattering islands, the lattice vector is also 3.58 nm. (d-f) Corresponding band structures calculated by the muffin-tin approximation. The band structures for the designs (b) and (c) reflect (nearly) separated s (blue) and p (orange) orbital bands. (g-i) The LDOS for these three designs; green for the on-site positions, purple for the*

*bridge positions between the sites. Blue arrows indicate the s orbital Dirac point, orange arrows indicate the p orbital flat band and the p orbital Dirac point. A broadening of 40 meV is included to account for scattering with the bulk.*

The theoretically designed honeycomb lattices are presented in Fig. 1, with the original lattice by Gomes *et al.* [16] (Fig. 1(a)), and two new designs (Fig. 1(b, c)). We have calculated the electronic band structure of these lattices by solving the Schrödinger equation with a muffin-tin potential accounting for the rosettes of CO molecules as repulsive scatterers. [17] The resulting band structures are presented in Fig. 1(d-f). In addition, we fitted the muffin-tin band structure with a tight-binding model based on artificial atomic sites in a honeycomb lattice; each atomic site has one *s* orbital and two in-plane *p* orbitals, and we assume *s-s*, *s-p* and *p-p* hopping between neighboring sites (see SM Section A). The tight-binding parameters have been adapted to obtain an optimal fit with the muffin-tin calculations (see Fig. S2). The calculations predict a single Dirac cone (blue color) for the lattice by Gomes *et al.* (Fig. 1(d, g)) in agreement with the experimental results reported. For this lattice, our calculations show that the next band (orange color) is strongly dispersive. In order to be able to separate the on-site *s-* and *p* orbitals we increased the on-site quantum confinement by using single and double CO rosettes as potential barriers. The design presented in Fig. 1(b) is based on single CO rosettes. [18] In this case, two dispersive *s* orbital bands emerge, forming a Dirac cone (blue). The four *p* orbital bands (orange) contain a (nearly) flat band and two dispersive bands forming a Dirac cone. However, in this design the *s* and *p* bands are not separated. In order to prevent this *s-p* hybridization, the on-site *s-* and *p* energy levels must be better separated by quantum confinement. This is achieved with the lattice presented in Fig. 1(c) (double-ringed CO rosettes as scatterers), showing the *p* orbital flat band and Dirac cone, well separated in energy from the lower *s* Dirac cone. The LDOS calculated for designs (b) and (c) (Fig. 1(h, i)) display a double peak with a minimum, reflecting the $s$ Dirac cone, followed by a single peak with high LDOS due to the $p$ orbital flat band, followed by a second double peak due to the $p$ orbital Dirac cone. This indicates that our lattices are appropriate electronic quantum simulators for the study of the in-plane $p$ orbital physics.

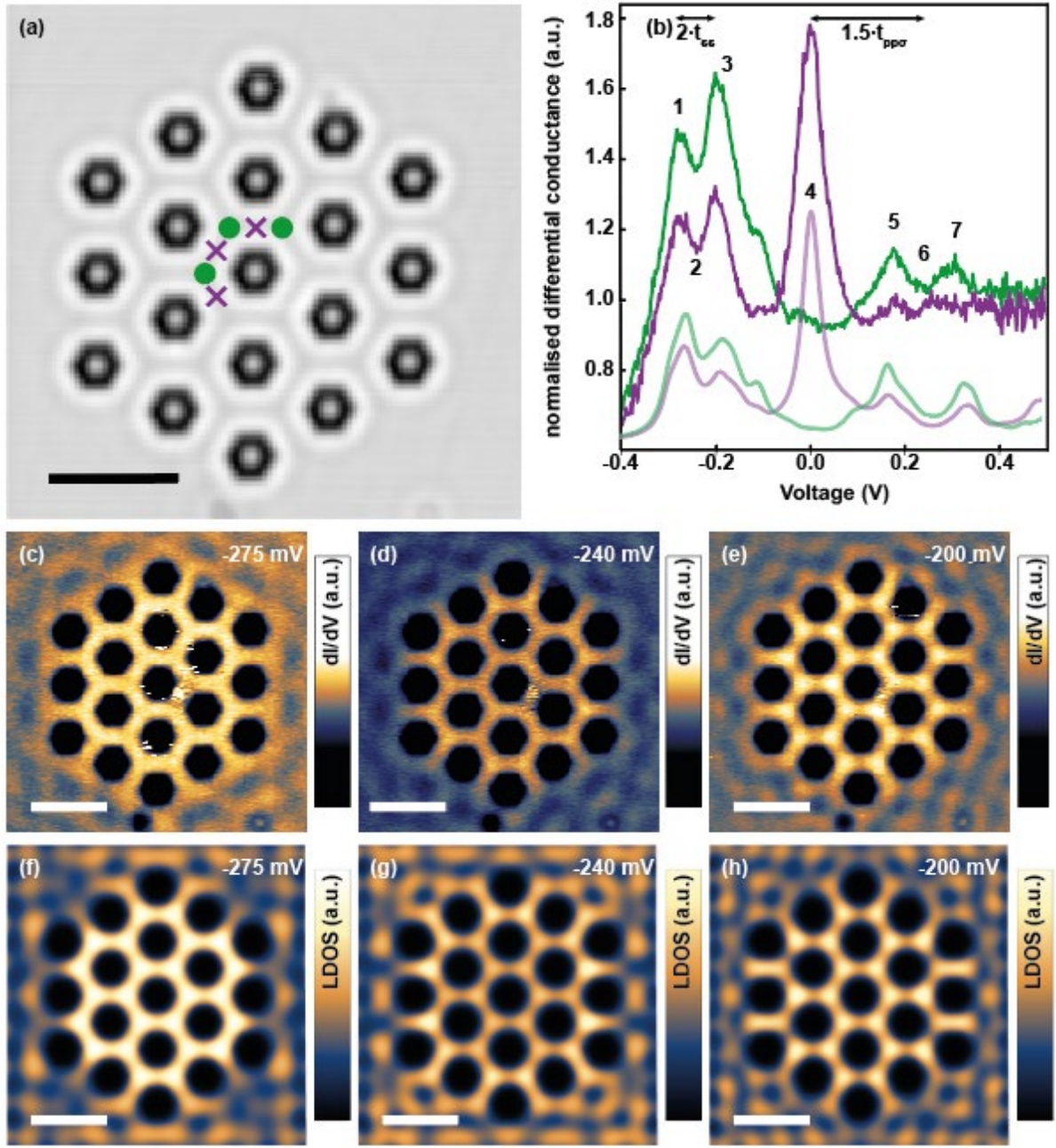

**Figure 2.** Scanning tunneling spectroscopy and electron probability maps of an artificial honeycomb lattice with separated s- and p bands.

*(a) Scanning tunneling microscopy image of the artificial honeycomb lattice prepared with double-ringed rosettes according to scheme 1(c); a detailed image for a similar lattice is presented in the Fig. S5.*

*(b) The LDOS, i.e. (dI/dV)$_{lattice}$/(dI/dV)$_{Cu}$, vs. bias voltage V, measured by scanning tunneling spectroscopy, on top of the artificial atom sites (green) and bridge sites (purple). The LDOS calculated using the muffin-tin approach is replotted in light green and light purple for comparison. The magnitudes of t$_{ss}$ and t$_{pp\sigma}$ are indicated.*

*(c, d, e) Spatially resolved LDOS maps in the energy region of the lowest Dirac cone (points 1-3 in Fig. 1(b)) measured at constant height with (f, g, h) the same maps calculated with a muffin-tin potential landscape. The high density of states at the sites reflect s orbital bands. Scale bars are 5 nm.*

First, we present an overall electronic characterization of the honeycomb lattice according to the design shown in Fig. 1(c). The results on the other lattice (Fig. 1(b)) are given in SM Section C. Fig. 2(a) shows a scanning tunneling microscope image using a Cu tip. Details are presented in Fig. S5, displaying a nearly identical lattice but now imaged with a CO-terminated tip. The LDOS could be probed with scanning tunneling spectroscopy by placing the metallic Cu-coated tip above the center of the artificial sites (green circles in Fig. 1(c) and 2(a)) and on bridge sites between the lattice sites (purple crosses); the bias voltage was changed over the entire voltage region of the Cu surface state between V = -0.4 and +0.5 V. The LDOS, i.e. normalized dI/dV vs. bias voltage, [16] spectra on the on-site and bridge site positions are presented in Fig. 2(b), see Fig. S6 for details; they should be compared with the theoretical muffin-tin spectra, for convenience replotted from Fig. 1(i) in light colours. The first double peak (peaks 1 and 3) corresponds to two s orbital bands forming a Dirac cone, the minimum indicates the Dirac point (point 2). The two maxima correspond to the high LDOS at the M points (see SM Section F); if the overlap integral between neighboring s orbitals is neglected, the distance between these two maxima provide a good estimation for two times the hopping term between the nearest-neighbor $s$ orbitals, i.e. $2t_{ss}$ (see SM Section F). The $t_{ss}$ value that we obtain is 45 meV. From a tight-binding fit, taking the overlap into account, we find 60 meV. The two $s$ orbital bands do not show the typical bonding (lowest $s$ band) and anti-bonding (higher $s$ band) character. An analytical tight-binding model presented in SM Section B, provides a detailed explanation.

Around V = 0 V, a very strong LDOS peak is observed on the bridge sites, while the LDOS on the lattice sites is very low (peak 4). A comparison with the muffin-tin band structure, and the tight-binding fit to it, reveals that this strong resonance localized between the sites is due to the flat band originating from $p$ orbitals. The high electron probability observed between the lattice sites will be discussed in detail below. Between 0.1 and 0.4 V, we find a second double peak with a minimum. Comparison with our calculations shows that this feature reflects the dispersive $p$ orbital bands; the minimum corresponds to the Dirac point (point 6), the lower maximum (peak 5) reflects the high LDOS at the M point. The maximum at higher energy (peak 7) corresponds to the third and fourth $p$ orbital bands. If the orbital overlap and residual $s$-$p$ hybridization are neglected, the energy difference between the flat band maximum and the Dirac point is 1.5 $t_{pp\sigma}$; from this, $t_{pp\sigma}$ is found to be 160 meV. From the muffin-tin calculations combined with a tight-binding fit we find a value of 127 meV (see Table S1).

Figures 2(c, d, e) display energy-resolved LDOS maps in the energy region of the $s$ bands measured over the entire lattice at a constant tip-sample distance, while the panels below (Fig. 2(f-h)) show the electron probabilities calculated with the muffin-tin model. There is a good agreement between the observed and calculated LDOS; the large on-site LDOS reflects the on-site $s$ orbitals, the LDOS at the Dirac point is much lower, but does not vanish completely. This reflects a certain broadening of the resonances due to the coupling of the lattice states with surface states outside the lattice and with Cu bulk states. A discussion of the LDOS maps in the *s*

band region from the tight-binding perspective is given in SM Section G.

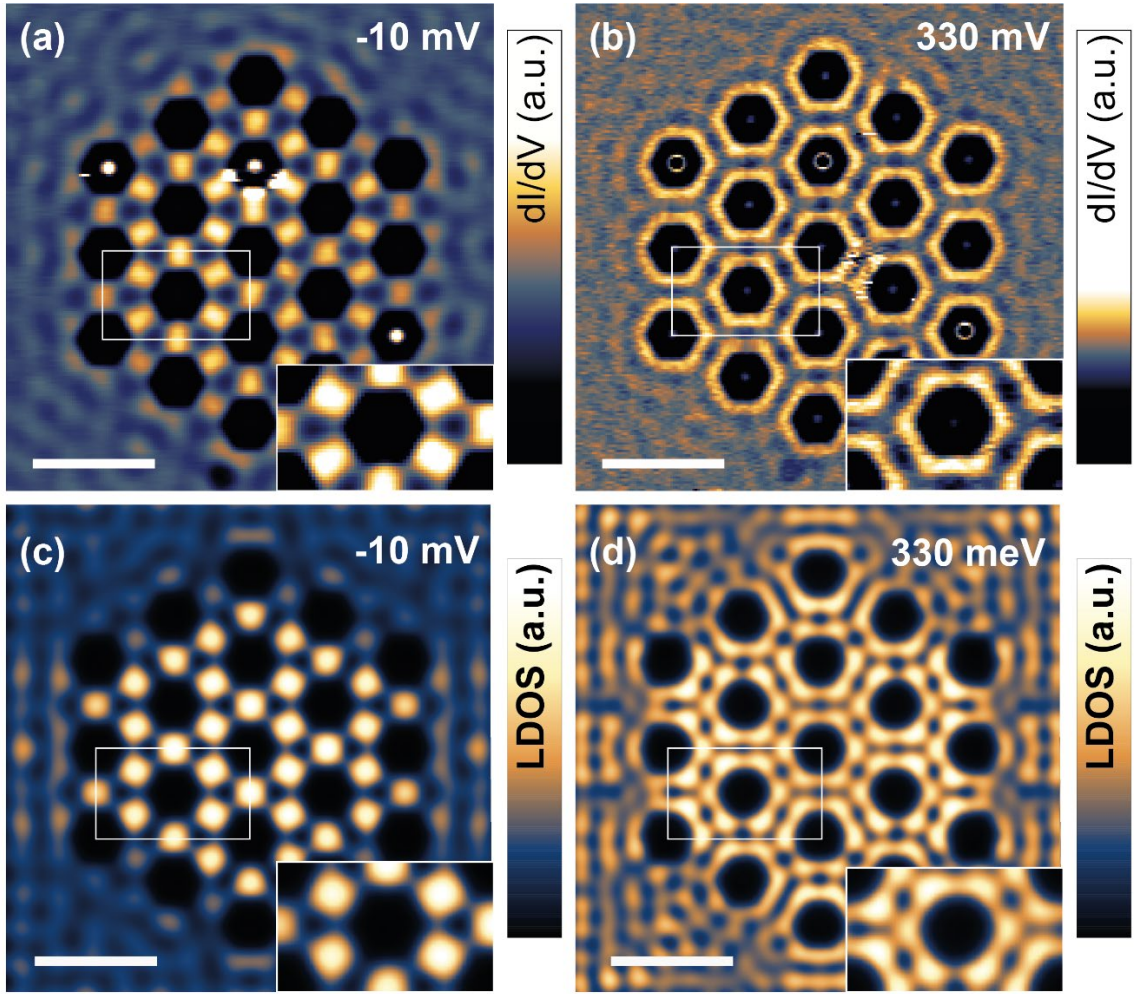

**Figure 3.** Electron probability (LDOS) maps in the energy region of the p orbital flat band and p

orbital Dirac cone obtained by energy-resolved scanning tunneling microscopy at constant height.

*Spatially resolved LDOS measured at (a) the flat band energy [point 4 in Fig. 2(b)] showing*

*patterns of very high electron probability at bridge sites, and very low probability on the atomic*

*sites; (b) in the energy region of the p orbital Dirac cone [point 7 in Fig. 2(b)].*

*The LDOS calculated using a muffin-tin approach for (c) the flat band showing a pattern of large electron densities between the sites, and very low electron density on the sites, to be compared with the experimental result in Fig. 3(a); (d) the energy region of the p orbital Dirac cone, showing a good agreement with the intriguing patterns experimentally observed. More information can be found in the SM. The inserts show a magnification. Scale bars are 5 nm.*

Maps of the electron probability measured in the energy region of the $p$ bands are presented in Figs. 3(a) and (b); Figs. 3(c) and (d) show the calculated results. The electron probability pattern at the flat-band energy is remarkable, with a very high electron probability between the sites, and a very low probability on the sites [Figs. 3(a), (c) and inserts]. In addition, the electron probability (LDOS) map in the region of the $p$ orbital Dirac cone show remarkable and detailed patterning [Figs. 3(b), (d) and insert], see also SM Section H. The low on-site electron probability on the center of the lattice sites show that these two bands are formed from $p$ orbitals.

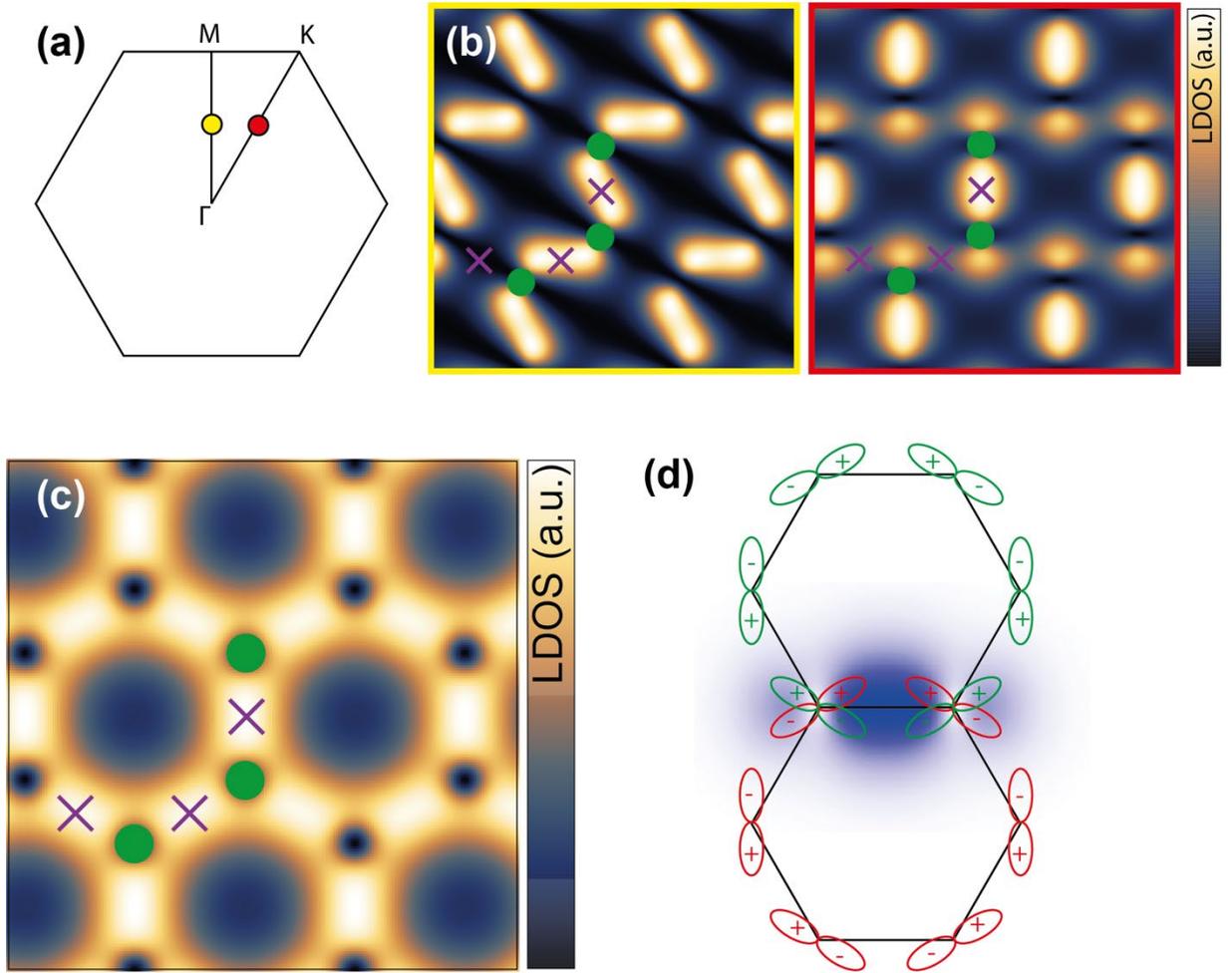

**Figure 4.** Tight-binding calculation of the interference patterns of the in-plane $p_x$, $p_y$ orbitals in the honeycomb geometry and the resulting electron probabilities in the p type flat band.

*(a) Scheme of the Brillouin zone with Γ, M and K points indicated. The yellow (red) circles denote the positions in the zone used in panel (b).*

*(b) Two spatial patterns due to interference of the $p_x$, $p_y$ orbitals in the honeycomb geometry at the flat band energy (-0.01 V) at the two points in the Brillouin zone indicated in panel (a). Artificial atom sites (green) and bridge sites (purple) are indicated.*

*(c) The overall electron probability at the flat band energy obtained from the interference patterns (see panel (b)) and summed over the entire Brillouin zone. Strong electron probabilities*

*are observed on bridge sites (purple crosses) as in the experimental maps.*

*(d) Representation of the electron probability map at the flat band energy by construction of Wannier-like eigenstates from p orbitals organized around a hexagon. The dark blue color indicates high electron probability. See also Fig. S9(c).*

The intricately patterned electron probabilities observed in the energy region of the *p* orbital bands in the honeycomb lattice require further investigation. The interaction of in-plane *p* orbitals at the sites of a honeycomb lattice can best be described as orbital interference by geometric frustration. [6] We have calculated these interference patterns by using the original tight-binding theory, [4, 6] see Figure 4. The results of the muffin-tin calculations combined with a tight-binding parameter fit are presented in Figs. S2(c) and (d). At the flat band energy, different points in the Brillouin zone show distinct interference patterns from the in-plane *p* orbitals, two of them being presented in Fig. 4(a, b). The overall sum of the electron probability patterns over the Brillouin zone at the energy of the flat band is presented in Fig. 4(c), showing a strongly enhanced electron probability on the bridge sites, in full agreement with the experimental results. Likewise, as originally proposed in Ref. 4, Wannier-like eigenstates with the flat band energy can be constructed around each CO rosette of artificial sites resulting in a high electron probability between the lattice sites (also called bridge sites), see Fig. 4(d). This remarkable spatial electron probability pattern in the flat band agrees with the experimental results and the result of muffin-tin calculations, see Fig. 3. In addition, a comparison between Fig. 3(b) and 3(d) shows that the spatial patterns of the LDOS in the *p* orbital Dirac region are well reproduced by the muffin-tin calculations.

Our results show that solid-state electronic honeycomb lattices can be designed in such a way that in-plane $p$ orbital physics fully emerges. The design is purely based on the lattice geometry and the degree of quantum confinement and inter-site coupling. These concepts can, therefore, be directly transferred to two-dimensional semiconductors in which the honeycomb geometry is lithographically patterned, [19-22] or, obtained by nanocrystal assembly. [7, 23] Such honeycomb semiconductors can be incorporated in transistor-type devices in which the Fermi level and thus the density of the electron gas can be fully controlled. [22, 24] For instance, a partial filling of the flat band can result in electronic Wigner crystals, new magnetic phases and superconductivity. [4, 6] Hence, we present a feasible geometric platform for real materials opening the gate to novel electronic quantum phases, both in the single-particle regime [7, 17, 25, 26] as in the regime with strong interactions. [27-29]

METHODS

The measurements were obtained in a Scienta Omicron LT-STM. It was operated at a base temperature of 4.5 K and with a pressure in the $10^{-10}$ mbar range. A clean Cu(111) surface was prepared by multiple sputtering and annealing cycles. [30] CO molecules were deposited on the sample placed in a cooled measurement head by leaking in gas at a pressure of $2 \cdot 10^{-10}$ mbar for 3 min. The STM tips were PtIr coated with Cu due to tip preparation. Atomic scale lateral manipulation of the CO molecules was performed to build the honeycomb lattices using previously obtained parameters of 40 nA and 10 mV. [16, 31, 32] Unless mentioned otherwise, all STM topography images were acquired at a constant current of 1 nA and 500 mV. Wave function mapping and differential conductance spectroscopy were performed using constant-height mode with a lock-in amplifier providing a 273 Hz bias modulation with an amplitude between 5 and 20 mV rms. Experimental data was analyzed with the SPM analysis software Gwyddion 2.49 and/or Python 3.7.

The design of the CO rosettes was determined by previously acquired knowledge about CO manipulation [16] and muffin-tin band structure calculations. The double-ringed rosette consists of 18 CO molecules arranged in two rings placed around a central (empty) Cu lattice site as shown in Fig. 1(c). This central site was left clear for ease of building. The rosettes were placed at a 3.58 nm spacing (14 Cu atomic sites) along close-packed Cu atomic rows.

All band structures and theoretical LDOS maps shown Fig. 1-3 of the main text were calculated using the muffin-tin model. The surface state of Cu(111) is modelled as a two-dimensional electron gas with an effective electron mass of 0.42 times the free electron mass, at a constant potential. The CO molecules are portrayed as discs with a diameter of 0.6 nm and a repulsive potential of 0.9 eV. These parameters were used previously to successfully describe the CO on copper system. [33] When CO molecules were placed close together and the radii overlapped, the potential of that area was added together and increased to 1.8 eV. The one electron Schrödinger equation was solved numerically for this system to determine the band structure (periodic case) and LDOS maps (finite size). For the LDOS maps, Neumann boundary conditions were applied. In order to obtain the maps shown, a broadening of 0.04 eV was included.

**Supporting Information**.

Extended theoretical background and experimental verification (PDF)

**Corresponding Author**

*D.Vanmaekelbergh@uu.nl

**Present Addresses**

† 1) Physical Measurement Laboratory, National Institute of Standards and Technology, Gaithersburg, Maryland 20899, USA

2) Department of Physics, Georgetown University, Washington, District of Columbia 20007, USA

**Author Contributions**

The manuscript was written through contributions of all authors. All authors have given approval to the final version of the manuscript. ‡These authors contributed equally.

**Funding Sources**

Financial support from the Foundation for Fundamental Research on Matter (FOM, grants 16PR3245 and DDC13), which is part of the Netherlands Organization for Scientific Research (NWO), as well as the European Research Council (Horizon 2020 "FIRSTSTEP", 692691) is gratefully acknowledged.

**Notes**

The authors declare no competing financial interests.

# Supporting Information for

## *p* Orbital flat band and Dirac cone in the electronic honeycomb lattice

Thomas S. Gardenier[1†], Jette J. van den Broeke[2†], Jesper R. Moes[1], Ingmar Swart[1], Christophe Delerue[3], Marlou R. Slot[1], Cristiane Morais Smith[2], Daniel Vanmaekelbergh[1*].

[1]Debye Institute for Nanomaterials Science, Utrecht University, The Netherlands.

[2]Institute for Theoretical Physics, Utrecht University, The Netherlands.

[3]Université de Lille, CNRS, Centrale Lille, Yncréa-ISEN, Université Polytechnique Hauts-de-France, UMR 8520 - IEMN, F-59000 Lille, France

Correspondence to: D.Vanmaekelbergh@uu.nl

**This PDF file includes:**

Supporting Information Sections A to K

Figs. S1 to S12

Table S I

## Section A.    Tight-binding analysis of the s and p orbital bands in artificial honeycomb lattices

In the main text, we have compared the experimental spectra and spatial LDOS maps with a muffin-tin calculation (see Fig. 2,3), showing a very good agreement between experiment and theory. It is however very insightful to also perform simple tight-binding (TB) calculations, in order to show which atomic site orbitals are involved in the band formation and to estimate the strength of the coupling between specific orbitals.

In the TB approximation, we assume that due to the repulsive potential of the CO rosettes atomic sites can be defined, with $s$ and $p$ orbitals (see Fig. S1(a)). We can choose the on-site energy of the $s$ and (two) $p$ energy levels, they are denoted as $e_s$ and $e_p$. The interaction energy, i.e. hopping (in eV) between the $s$ orbitals of two neighboring sites is denoted by $t_{ss}$, the hopping between $s$ and in-plane $p_x$ and $p_y$ orbitals by $t_{sp}$, the $\sigma$ type interaction integral between the in-plane $p$ orbitals on adjacent sites by $t_{pp\sigma}$, and the $\pi$ hopping between in-plane $p$ orbitals by $t_{pp\pi}$. These hoppings are depicted in Fig. S1(a). We neglect the on-site orbitals at higher energy in this simple approximation. The TB Hamiltonian is:

$$\begin{pmatrix} H_1 & H_2 \\ H_2^\dagger & H_1 \end{pmatrix} \text{ with, } H_1 = \begin{pmatrix} e_s & 0 & 0 \\ 0 & e_p & 0 \\ 0 & 0 & e_p \end{pmatrix} \text{ and}$$

$$H_2 = \begin{pmatrix} t_{ss}\left(e^{\frac{ik_x}{\sqrt{3}}} + 2e^{-\frac{ik_x}{2\sqrt{3}}}\cos\left[\frac{k_y}{2}\right]\right) & t_{sp}\left(-e^{\frac{ik_x}{\sqrt{3}}} + e^{-\frac{ik_x}{2\sqrt{3}}}\cos\left[\frac{k_y}{2}\right]\right) & t_{sp}i\sqrt{3}e^{-\frac{ik_x}{2\sqrt{3}}}\sin\left(\frac{k_y}{2}\right) \\ t_{sp}\left(e^{\frac{ik_x}{\sqrt{3}}} - e^{-\frac{ik_x}{2\sqrt{3}}}\cos\left[\frac{k_y}{2}\right]\right) & -t_{pp\sigma}e^{\frac{ik_x}{\sqrt{3}}} - \left(\frac{t_{pp\sigma}}{2} - \frac{3t_{pp\pi}}{2}\right)e^{-\frac{ik_x}{2\sqrt{3}}}\cos\left(\frac{k_y}{2}\right) & -(t_{pp\sigma} + t_{pp\pi})\frac{i\sqrt{3}}{2}e^{-\frac{ik_x}{2\sqrt{3}}}\sin\left(\frac{k_y}{2}\right) \\ -t_{sp}i\sqrt{3}e^{-\frac{ik_x}{2\sqrt{3}}}\sin\left(\frac{k_y}{2}\right) & -(t_{pp\sigma} + t_{pp\pi})\frac{i\sqrt{3}}{2}e^{-\frac{ik_x}{2\sqrt{3}}}\sin\left(\frac{k_y}{2}\right) & t_{pp\pi}e^{\frac{ik_x}{\sqrt{3}}} - \left(\frac{3t_{pp\sigma}}{2} - \frac{t_{pp\pi}}{2}\right)e^{-\frac{ik_x}{2\sqrt{3}}}\cos\left(\frac{k_y}{2}\right) \end{pmatrix}$$

In Fig. S1(b-f), we show how different hopping parameters influence the band structure. Here, we neglect the overlap integrals. If $t_{sp}$ is zero, there is no hopping between the $s$ and $p$ orbitals, thus no hybridization, and the bands formed should have pure $s$ character (two bands) and pure $p$ character, (four bands). The $s$ bands form a Dirac cone with the Dirac point at zero energy. The orthogonal in-plane $p$ orbitals are not commensurable with the trigonal binding structure (see Fig 4). This results in two flat bands, with a Dirac cone between these flat bands (Fig. S1(b)). If $t_{pp\pi}$ is non-zero, the two flat bands acquire a dispersion, while the $p$ orbital Dirac point is preserved. Here we also take $t_{pp\pi}$ equal to $t_{pp\sigma}$, resulting in a fourfold degeneracy of the $p$ bands at the $\Gamma$ point, see Fig. S1(c).

In Fig. S1(d, e) we show what happens if the energy difference between the on-site $p$ and $s$ orbitals is lowered and if $s$-$p$ hopping is allowed. First, in panel (d), we show the bands with reduced on-site energy difference, but still with $t_{sp}$ being zero. This results in unrealistic crossing points between the $s$ and $p$ bands. The introduction of hopping between $s$ and $p$ orbitals of adjacent sites results in a grouping of three lower bands and three higher bands, separated by a gap. There is a downwards shift of the *lower* Dirac cone, and an upwards shift of the *second* Dirac cone. The lower flat band touches the lower Dirac cone. Finally, panel (f) shows an example if all hopping parameters are non-zero; the two Dirac cones are preserved, but the originally flat bands obtain a dispersion due to $\pi$ hopping.

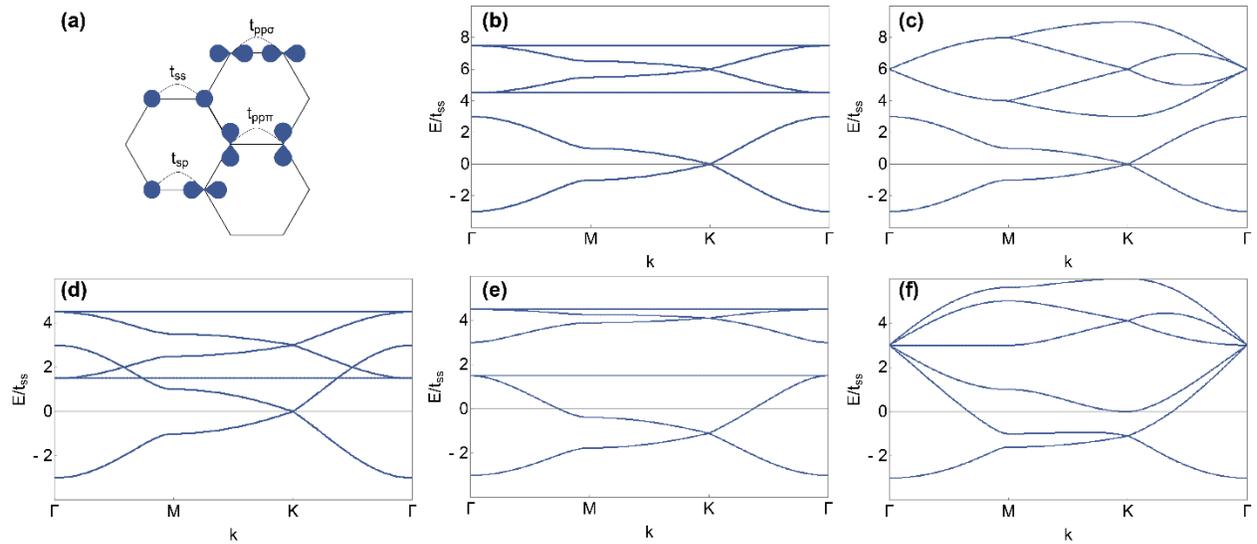

**Figure S1.** Tight-binding band structures, for various values of the hopping parameters.

(a) Scheme of the possible hoppings between $s$ and in-plane $p$ orbitals of adjacent atomic sites in the honeycomb lattice.

(b) A generic band structure for pure $s$ and $p$ orbital bands, by assuming there is no coupling between $s$ and $p$ orbitals, i.e. $t_{sp} = 0$ and no $\pi$ hopping.

(c) A band structure showing the effect of $\pi$ hopping between the in-plane $p$ orbitals, resulting in some dispersion of the bands that were flat in B.

(d-f) The effect of reducing the energy difference between the on-site $s$ and $p$ energy levels.

(d) Bands cross as $t_{sp}$ is set equal to zero.

(e) The effect of $s$-$p$ coupling results in two groups of ($sp_2$) bands, with Dirac cones and flat bands.

(f) When $t_{pp\pi}$ is set to non-zero, the formerly flat bands obtain a dispersion.

**Section B.      Parametrization of the tight-binding hoppings in order to obtain maximum agreement with the experimental results and the muffin-tin approximation.**

In the main text, the experimental results are compared with a muffin-tin calculation of the band structure. The potential landscape of the individual CO molecules and CO rosettes on Cu(111) can be modelled using a muffin-tin (MT) potential. This is done by adding disk-shaped potential barriers to an otherwise flat potential landscape, resulting in an upside-down muffin tin like structure. In this work, disk diameters of 0.6 nm and potential heights of 0.9 eV were used to account for each CO. By analytically Fourier-transforming the muffin-tin potential landscape and using Bloch-type wave functions, we calculate the electronic band structure for electrons in the honeycomb lattices presented in Fig. 1.

In order to be able to discuss the strength of the hoppings between the on-site $s$ and $p$ orbitals, we have varied the tight-binding (TB) hoppings and on-site energies in order to obtain the best agreement between the MT band structure (in agreement with experimental results) and the TB approximation. Here, we have also accounted for the overlap integrals in the TB calculation; orbital overlap between the $s$ orbitals is denoted as $s_{ss}$, between the $s$ and in-plane $p$ orbitals as $s_{sp}$, and between in-plane $p$ orbitals as $s_{pp\sigma}$ and $s_{pp\pi}$.

We have varied the TB parameters such that the MT and TB band structures agree as well as possible. In finding the best agreement, we focus on the lower bands, and allow for differences between TB and MT results for the higher $p$ orbital bands. The MT and TB band structures and the corresponding designs are shown in Fig. S2. It can be seen that the $s$ and $p$ orbital bands of the experimentally studied lattices can be well approximated with a TB model with $s$ and $p$ orbitals only, except for the highest $p$ band. The corresponding parameters are given in Table S1. Because there are 10 fitting parameters, this fit might not be unique. We would like to remark that the relative values of the main hopping parameters $t_{ss}$ and $t_{\sigma pp}$ seem to be very reasonable, seen from a chemical orbital perspective. We show calculations for a lattice similar to the one studied previously by Gomes *et al.,* [15] the two lattices that we have examined, and a lattice with a triple-ringed CO rosette. When the rosettes are enlarged, on-site quantum confinement increases the energy difference between the on-site $s$ and $p$ energy levels.

The increasing agreement between TB and MT with increasing confinement has several origins. First, orbitals higher than $p$ are not incorporated in the TB model. The influence of these orbitals on the lower bands is not completely neglectable and is automatically taken into account in the MT calculations, but not in the TB calculations. Thus, the simple TB approximation becomes more accurate when the energy difference between the on-site energy levels increases. In addition, the $s$ and $p$ orbital bands become more pure when the on-site energy separation between the $s$ levels and $p$ levels increases. We were able to design artificial lattices that unambiguously show two separated Dirac cones and a flat band.

A second factor that improves the TB approximation is that for increasingly larger rosettes, the influence of the orbital overlap and $t_{pp\pi}$ hopping decreases. When $t_{pp\pi}$ becomes neglectable, the lowest and highest $p$ orbital bands lose their dispersion and become genuine flat bands.

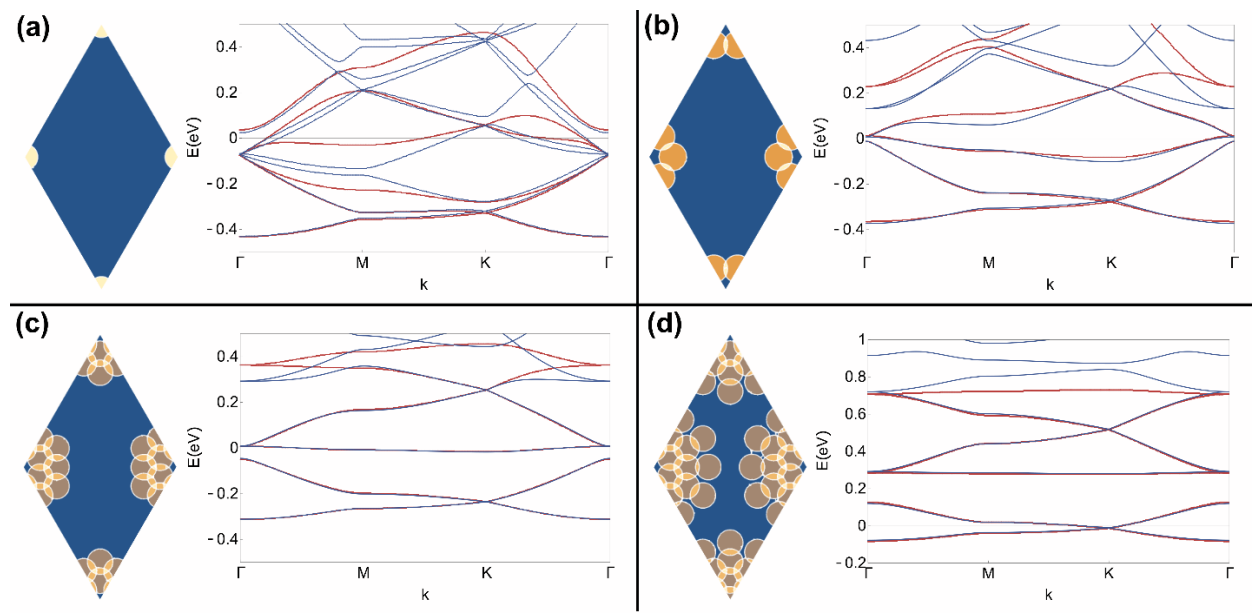

**Figure S2.** Fitting of the calculated muffin-tin band structures with a tight-binding model.

Each panel contains a unit cell (blue diamond) of the corresponding lattice with CO molecules portrayed as brown discs. The corresponding band structures are shown next to it with the muffin-tin results in blue and tight-binding results in red. The tight-binding hopping and overlap parameters are presented in Table SI.

(a) The lattice reported by Gomes *et al.* [1] with a single CO molecule as scatterer.

(b) A lattice with single-ringed CO rosettes as scatterers.

(c) A lattice with double-ringed CO rosettes as scatterers.

(d) A lattice with triple-ringed CO rosettes (not experimentally studied).

**Table S I**

Fitting parameters for the designs shown in Fig. S3. Units, where applicable are eV.

| | $t_{ss}$ | $t_{sp}$ | $t_{pp\sigma}$ | $t_{pp\pi}$ | $e_s$ | $e_p$ | $s_{ss}$ | $s_{sp}$ | $s_{pp\sigma}$ | $s_{pp\pi}$ |
|---|---|---|---|---|---|---|---|---|---|---|
| single CO | -0.09 | -0.09 | -0.11 | -0.11 | -0.24 | -0.075 | 0.06 | 0.06 | 0.15 | 0.15 |
| single-ringed rosette | -0.07 | -0.09 | -0.105 | -0.045 | -0.22 | 0.105 | 0.06 | 0.07 | 0.2 | 0.1 |
| double-ringed rosette | -0.062 | -0.06 | -0.1265 | -0.00825 | -0.22 | 0.185 | 0.1 | 0.1 | 0.05 | 0.05 |
| triple-ringed rosette | -0.034 | -0.05 | -0.131 | 0 | 0.01 | 0.49 | 0.04 | 0.01 | 0.03 | 0.01 |

## Section C.    Results obtained on an artificial honeycomb lattice formed by single-ringed CO rosettes

Figure S3 presents dI/dV vs. V spectra taken on the single-ringed CO rosette lattice, the design is shown in Fig. 1(b). We demonstrate the effect of the two normalization techniques shown in Fig. S6: subtraction (panel (b)) and division (panel (c)). In panel (a), in orange, we show an averaged spectrum taken on Cu(111), notice that this spectrum shows an increase in intensity above -0.2 V. This feature is also visible in the spectra taken on the atomic lattice sites (green) and bridge sites (purple). Subtraction of the dI/dV of the Cu(111) background partially corrects for this, but it is possible that features in the lattice LDOS remain clouded above 0.1 V.

In the region below 0.2 V, the bare and normalized spectra obtained on this lattice show clear features corresponding to the LDOS of the artificial lattice. The two peaks at -330 and -210 mV are assigned to the *s* orbital Dirac cone, more specifically to the M points around the Dirac point at -290 mV at K (see also main text). The strong feature at -0.1 V measured at the bridge sites reflects the *p* orbital flat band. The results are similar to those obtained with a lattice created with double-ringed CO rosettes. The large dip in intensity at -0.4 V corresponds to a normalization artefact and has no relevance for the band structure.

The spatial distribution of the LDOS over the lattice is presented in Fig. S4. At -330 mV, the LDOS intensity is strong on the lattice sites. The LDOS intensity is minimal at the Dirac point at -290 mV. At the second peak of the Dirac cone, at -210 mV, the intensity is high again on the lattice sites. The muffin-tin calculations reproduce the experimental maps well.

Figure S4 shows that there is a strong resonance peak at -90 mV for the bridge sites. This peak is absent on the atomic sites. Comparison with our muffin-tin and tight-binding calculations show that this peak reflects the *p* orbital flat band (see main text). The spatial distribution of the LDOS shows the remarkably strong intensity of the LDOS between the atomic sites, in full agreement with the results obtained on the other artificial lattice presented in the main text.

Although the spectra are not very different around 70 mV, we observe patterning throughout the lattice in the LDOS maps. This pattern corresponds to measurements taken on the double-ringed CO rosette lattice.

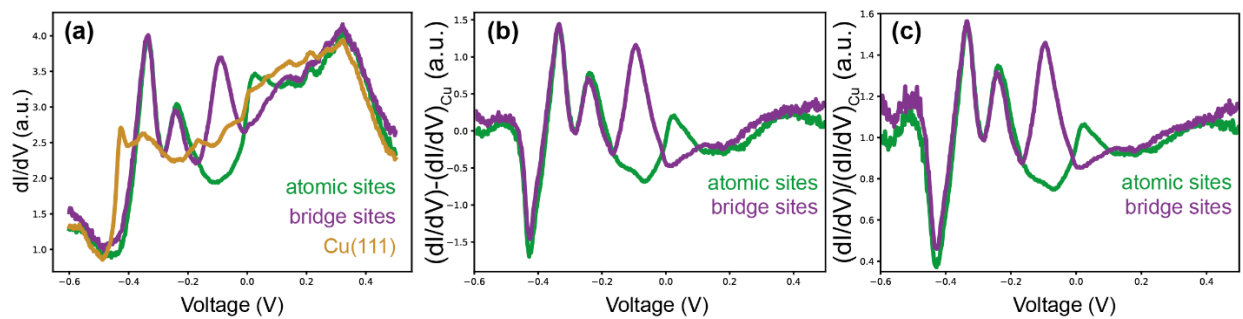

**Figure S3.** Effect of normalizing the raw spectra on an artificial honeycomb lattice formed by single-ringed CO rosettes by various techniques.

The (dI/dV) vs. V spectra were acquired on the lattice presented in Fig. 1(b), formed by an anti-dot lattice of single-ringed CO rosettes.

(a) Averaged spectra taken on two different symmetry positions in the lattice; on the centre of the atomic lattice sites (green) and at bridge sites in between (purple). The surface state measured on bare Cu(111) is shown in orange.

(b) The same spectra, but with the Cu(111) dI/dV background subtracted.

(c) The same spectra but divided by the Cu(111) dI/dV.

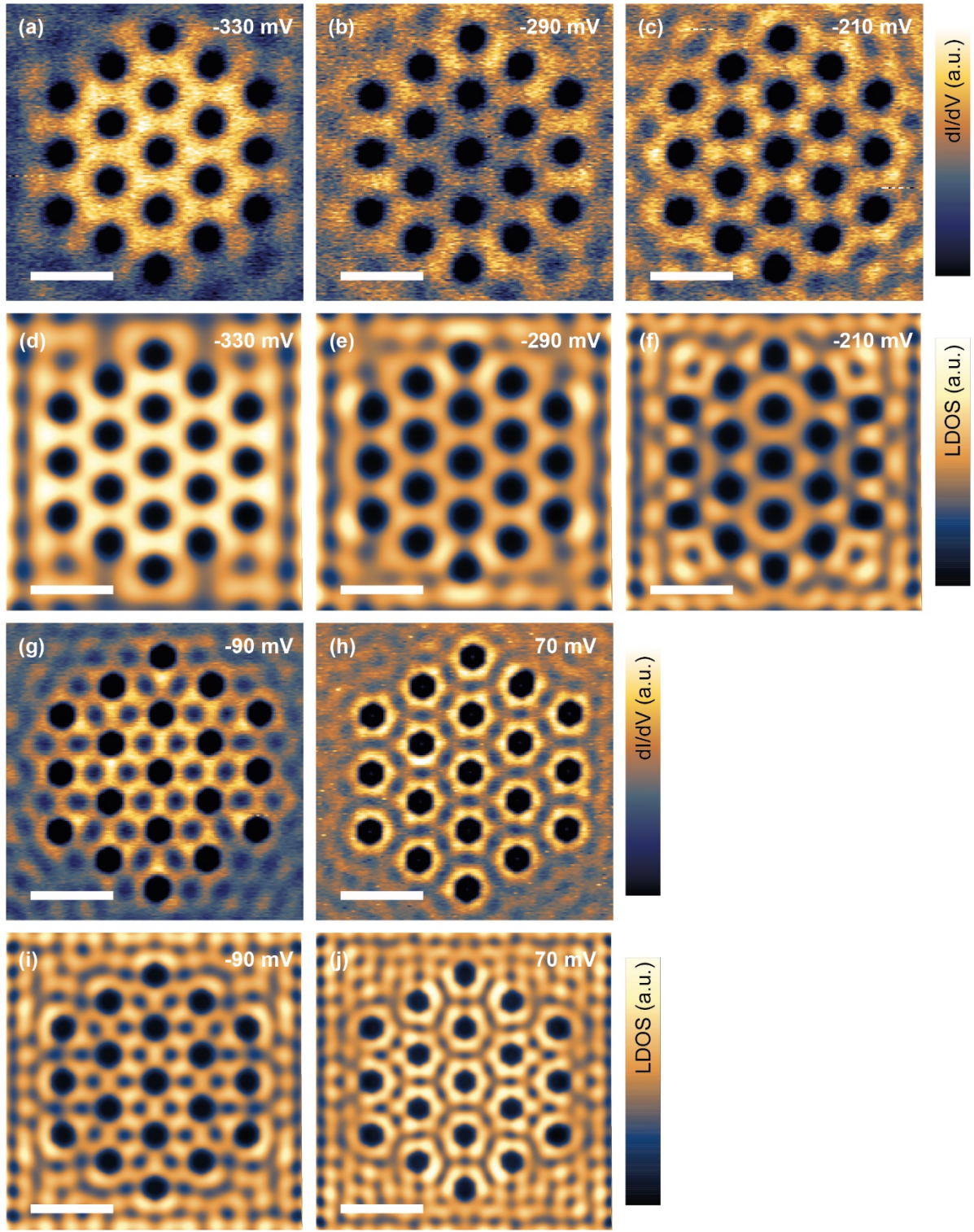

**Figure S4.** Electron probability maps obtained on an artificial honeycomb lattice formed by single-ringed CO rosettes.

(a-c) LDOS maps corresponding to the *s* orbital Dirac cone taken at the first maximum at -330 mV, the Dirac point at -290 mV and the second maximum at -210 mV. The maxima reflect the M points of the Dirac cone.

(d-f) Corresponding muffin-tin calculations for the electron probability corresponding to the maps in (a-c).

(g) LDOS map corresponding to the *p* orbital flat band at -90 mV, showing zero intensity on the lattice sites and very strong intensity in between the sites.

(h) LDOS map corresponding to higher energy *p* orbital bands at +70 mV.

(i, j) Corresponding muffin-tin calculations for the electron probability corresponding to the maps in (g, h), respectively. Scale bars are 5 nm.

# Section D.       The structure of an artificial honeycomb lattice created by rosettes of CO scatterers imaged with a CO tip.

The artificial honeycomb lattices studied in this work are prepared by creating a potential energy landscape to force the electrons of the Cu(111) surface state into a honeycomb geometry. The potential energy landscape is obtained by placing repulsive CO molecules acting as scatterers in rosettes, e.g. see Fig. 2. Fig. S5 presents a specific lattice imaged with a CO tip, allowing us to discern the individual CO molecules (absorbed on top of Cu atoms) as circular protrusions in each rosette, and even misplaced CO molecules. Please, also notice that we have placed CO scatterers around the lattice to isolate the lattice from the rest of the Cu(111) surface state.

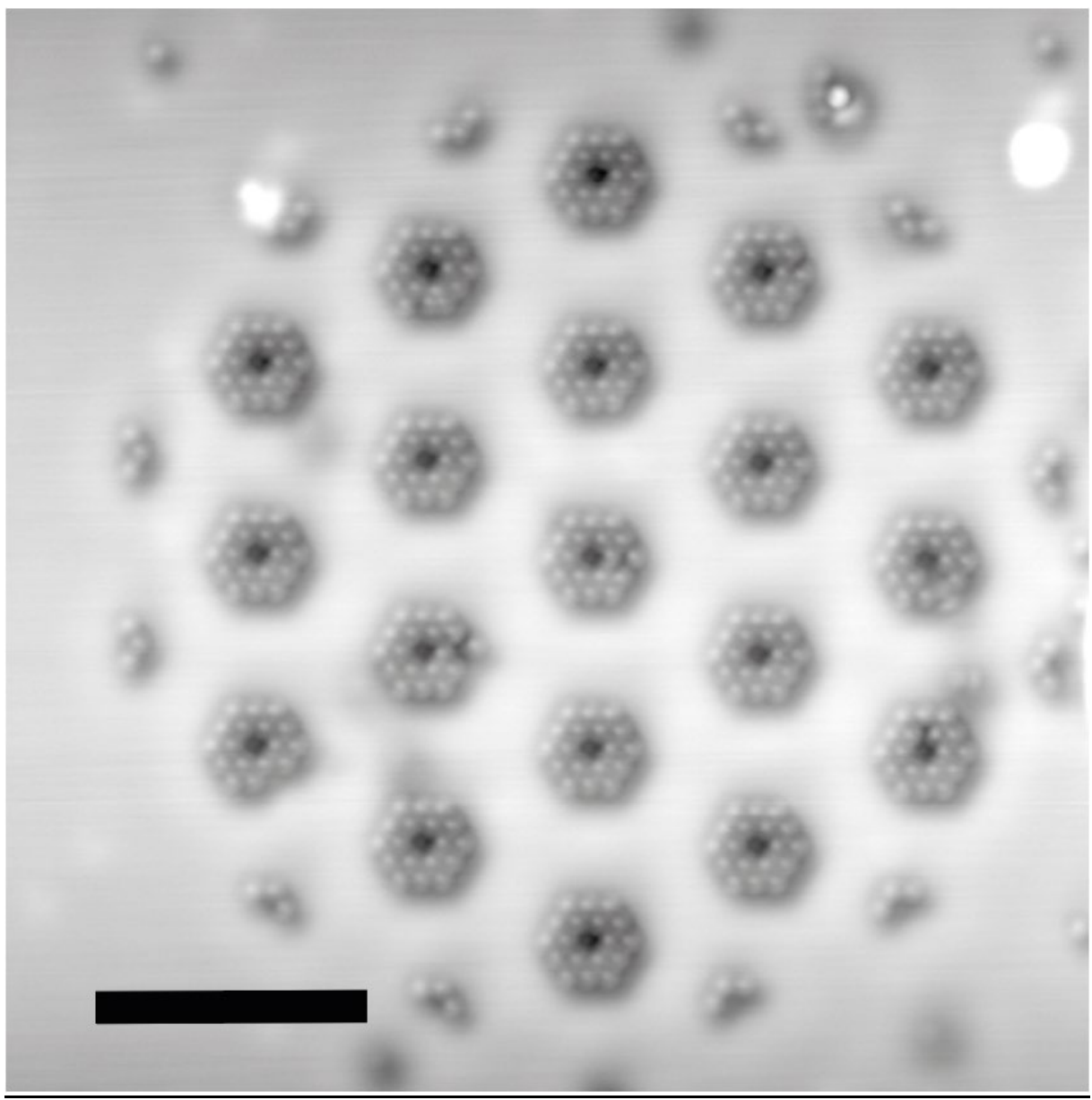

**Figure S5.** Detailed structure of the CO rosettes used to prepare artificial electronic lattices.

Constant current STM image of a honeycomb lattice with a lattice vector of 3.86 nm. The potential landscape is obtained by using double-ringed CO rosettes as repulsive scatterers for the surface state of the underlying Cu(111) surface. The purpose is to form artificial atomic sites located between the repulsive rosettes. This image was obtained with a CO-terminated tip. Each double-ringed rosette consists of 18 CO molecules, which are imaged as circular protrusions. Several defects or misplaced CO molecules can be spotted. Scale bar is 5 nm.

**Section E.     Effect of normalizing the raw spectra by various techniques**

In order to correct for the effects from the Cu sample and Cu tip, all dI/dV vs. V spectra in this manuscript have been presented as normalized spectra. This was done following a method used by Gomes *et al.* [1] The raw spectra were divided by an averaged dI/dV obtained on a bare Cu(111) surface, acquired with exactly the same settings and the same tip. This procedure should remove LDOS components of the tip and the Cu(111) sample. The normalization technique is the same as shown in Section 3. For the two different tip states and lattices the normalization provides reproducible results, a good indication that our normalization technique is sound.

In Fig. S6 we demonstrate the effect of two different normalization techniques. In panel (a) we show the raw spectra taken on bare Cu(111), and on lattice sites and bridge sites. First, the effect of quantum confinement in the lattice can be seen by the onset of resonances at higher energy than the onset of the bare surface state. Second, one can already see the peaks and valleys of interest in the spectra taken on the two positions in the lattice. However, the spectral intensities of the lattice should be corrected for the background related to substrate and tip. In panel (b), we subtracted the dI/dV of the Cu(111); a horizontal line through zero would form a reference. In panel (c) we divided the raw spectra by the Cu(111) background spectra; thus a horizontal line through 1 would now form a reference. The spurious peak at 0 V (green line) is absent in both cases. The procedure shown in (c) is the procedure used to represent the LDOS in the main text.

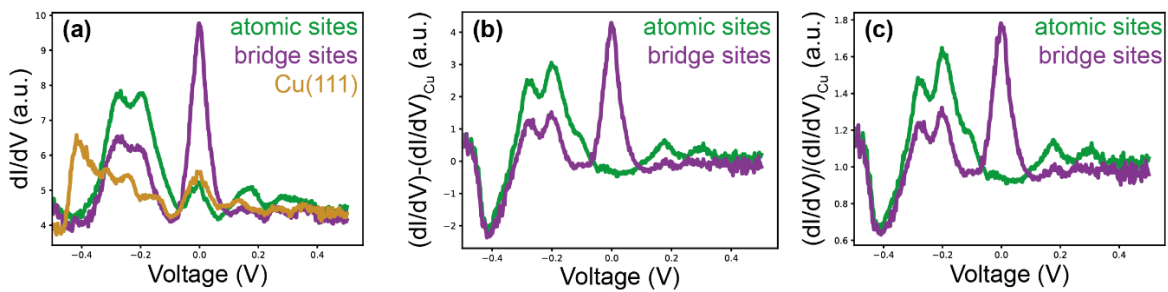

**Figure S6.** Effect of normalizing the raw spectra by various techniques.

The (dI/dV) vs. V spectra were acquired on the lattice presented in Fig. 2(a), formed by an anti-dot lattice of double-ringed CO rosettes.

(a) Averaged spectra taken on two different symmetry positions in the lattice; on the center of the atomic lattice sites (green) and at bridge sites in between (purple). The surface state measured on bare Cu(111) is shown in orange.

(b) The same spectra, but with the Cu(111) dI/dV background subtracted.

(c) The same spectra but divided by the Cu(111) dI/dV.

**Section F.      Analytic tight-binding model for the electronic honeycomb lattice in the absence of hybridization, broadening and orbital overlap.**

**DOS and spatially-resolved LDOS maps in the *s* sector**

We have calculated the local density of states $LDOS(E, \boldsymbol{r})$ in the absence of broadening, at energy $E$ and position $\boldsymbol{r}$ using

$$LDOS(E, \boldsymbol{r}) = \sum_{\boldsymbol{k}, \alpha} |\Psi_{\boldsymbol{k}, \alpha}(\boldsymbol{r})|^2 \, \delta(E - E_{\boldsymbol{k}, \alpha})$$

where $\Psi_{\boldsymbol{k}, \alpha}(\boldsymbol{r})$ is a state in band $\alpha$, with lattice momentum $\boldsymbol{k}$, and energy $E_{\boldsymbol{k}, \alpha}$. Due to the presence of the delta function, the $LDOS(E, \boldsymbol{r})$ can be rewritten as an integral over a constant-energy path in the Brillouin zone of the honeycomb lattice.

In the absence of $s - p$ hybridization, the energy bands in the *s* sector are given by [2]

$$E_{\boldsymbol{k}, \pm} = \pm t_{ss} |u(\boldsymbol{k})|,$$

where the zero of energy is set at the Dirac point, $t_{ss}$ is the hopping between nearest-neighbor *s* orbitals, and $u(\boldsymbol{k})$ is defined as $\sum_{j=1}^{3} \exp(i\boldsymbol{k}.\boldsymbol{\delta}_j)$, with the vectors $\boldsymbol{\delta}_j$ pointing to nearest neighbors of a site. By symmetry, the weight of each eigenstate $\Psi_{\boldsymbol{k}, \alpha}(\boldsymbol{r})$ is the same on the two Bloch waves formed by the *s* orbitals on the sub-lattices A and B. However, the phase between the two contributions differs by the angle $\theta_{\boldsymbol{k}} = \text{Arg } u(\boldsymbol{k})$. Since $E_{\boldsymbol{k}, \pm}$ does not depend on $\theta_{\boldsymbol{k}}$, the phase can be seen as a pseudo-spin degree of freedom. [3]

Figure S7(a) depicts constant-energy paths in the Brillouin zone. Close to zero energy, since the allowed states form cones around (K, K'), the paths consists of tiny (blue) circles, as shown at $E = \pm 0.2 t_{ss}$. The radius of the circles goes to zero at the Dirac point where the DOS vanishes. At increasing energy from the Dirac point, the constant-energy path tends to deviate from the circular shape. For $E = \pm t_{ss}$, the constant-energy path becomes a hexagon that touches the edge of the Brillouin zone at the M points. This leads to a maximum of the LDOS at these energies. In this simple model, in which hybridization with *p* orbitals is excluded and the overlap integral is neglected, the distance in energy between the two peak maxima in the LDOS(E) plot of the *s* bands is equal to $2|t_{ss}|$. This is used in the main text for a first estimation of $|t_{ss}|$. This value can be compared with the value obtained from a tight-binding model by fitting the 6 lowest bands (*s* and *p* bands), including *s-p* hybridization, and taking the overlap integrals into account.

Spatially-resolved LDOS maps are obtained by integration of $|\Psi_{\boldsymbol{k}, \alpha}(\boldsymbol{r})|^2$ on the constant-energy paths (Fig. S7(a)). In a tight-binding representation with one *s* orbital $\varphi_s(\boldsymbol{r} - \boldsymbol{R}_i)$ on each site $\boldsymbol{R}_i$, the LDOS for any allowed energy $E$ close to the Dirac point is just the superposition of the squared *s* orbitals, i.e., $\sum_i |\varphi_s(\boldsymbol{r} - \boldsymbol{R}_i)|^2$. The extra terms in $|\Psi_{\boldsymbol{k}, \alpha}(\boldsymbol{r})|^2$, which come from the cross terms between nearest-neighbor *s* orbitals, cancel out after integration over constant-

energy paths because they are proportional to $\cos(\theta_{\boldsymbol{k}})$. This is one explanation for the experimental results presented in Fig. 2: the two peaks reflecting the high density of states at the M points of the s orbital Dirac cone are nearly symmetrical in intensity, when measured on atomic sites (green curve in Fig. 2(b)) and on bridge sites (purple curve in Fig. 2(b)). This also explains the experimental LDOS maps of Fig. 2(c) and (e), with high intensities on the atomic sites and weaker intensity between the sites. However, in Fig. S8 we do not see the same symmetry effect. This can be understood as a consequence of the contribution of the energies far away from the Dirac point, where the approximation explained above is not valid. The symmetric density of states on the bridge sites can alternatively be explained by the influence of s-p hybridization in the highest s orbital that is not taken into account in Fig. S8.

## DOS and spatially-resolved LDOS maps in the *p* sector

The four energy bands (Fig. S7(b)) in a pure $p_x$, $p_y$ model (no s-p hybridization) with negligible π coupling are given by

$$E_{\boldsymbol{k},1} = -\frac{3}{2} t_{pp\sigma} \qquad E_{\boldsymbol{k},4} = -E_{\boldsymbol{k},1}$$

$$E_{\boldsymbol{k},2} = -\frac{1}{2} t_{pp\sigma} |u(\boldsymbol{k})| \quad E_{\boldsymbol{k},3} = -E_{\boldsymbol{k},2}$$

where $t_{pp\sigma}$ is the hopping term of σ type between nearest-neighbor p orbitals. [4, 5] The second and third bands have the same dispersion as the s orbital bands, provided that $t_{ss}$ is replaced by $t_{pp\sigma}/2$. The description with constant-energy paths, shown above, remains valid after this substitution. In particular, the DOS vanishes at zero energy (Dirac point) and presents a maximum at $E = \pm t_{pp\sigma}/2$, when the constant-energy paths form a hexagon connecting the M points of the Brillouin zone. The first and fourth bands in this $p_x$, $p_y$ model are totally flat, giving rise to the DOS in the form of Dirac delta functions at $E = \pm 3t_{pp\sigma}/2$ in absence of extra sources of broadening or dispersion (absence of π bonding or s-p hybridization).

Comparison between the band structure for the pure $p_x$, $p_y$ model (Fig. S7(b)) and the band structure calculated by solving the Schrödinger equation with a muffin-tin potential (Fig. 1(f)) shows that the lowest flat band and the p orbital Dirac cone are well distinguishable in the muffin-tin results and in the experimental LDOS spectra. The main differences appear in the upper part of the band structure due to strong coupling with higher-energy orbitals. This leads to a down shift of the dispersive band $E_{k,3}$, especially at the Γ point. In addition, the upper band $E_{k,4}$ is not flat anymore. It is thus wise to use the lowest bands to estimate the value of $t_{pp\sigma}$ from the experimental LDOS results, Fig. S7(b) and the values given in the main text.

Whereas the LDOS maps close to the Dirac point do not depend on energy in the s orbital model, the situation is totally different in the p orbital model due to the orbital degree of freedom. [4, 5] The orbital configuration on each site strongly varies with $\boldsymbol{k}$, explaining the remarkable patterns that were observed and presented in Fig. 3. In the case of the flat bands, the eigenstates $\Psi_{\boldsymbol{k},o}(\boldsymbol{r})$ can be written either in terms of Bloch states or, alternatively, as a linear superposition of Wannier-like localized states, which are all degenerate. One localized state exists per hexagonal plaquette. [4, 5] The configuration in terms of $p$ orbitals for the lowest flat band is depicted in Fig. S7(c) (see also Fig. 4). One $p$ orbital is tangential to the hexagon; as a consequence, the other $p$ orbital on the same site is then parallel to the bond external to the hexagon. The flatness of the band is explained by the cancellation of hopping terms to neighboring loops (interference effect) in absence of $\pi$ bonding. In this configuration, the LDOS map on nearest-neighbor atoms A and B is given by the squared amplitude of the two localized eigenstates that share this bond. Fig. S7(c) presents the summed amplitude calculated using a $p_x$ orbital of the form $x \cdot e^{\frac{-|r|}{\gamma}}$, where $\gamma = 0.25a$, $a$ being the lattice vector (same definition for $p_y$). It can be seen that the LDOS amplitude is very high at the center between two adjacent sites of the hexagon, in agreement with the experimental results and the muffin-tin calculation, see main text.

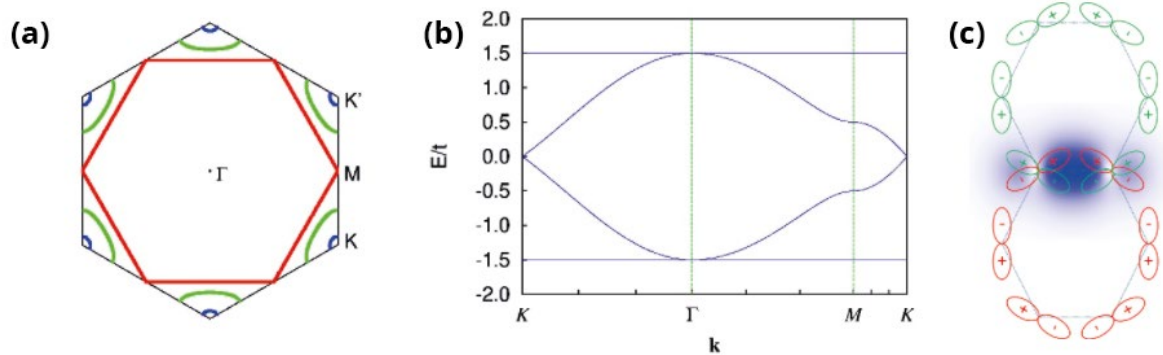

**Figure S7.** Tight-binding analysis without hybridization.

(a) Constant-energy paths in the Brillouin zone of the honeycomb lattice for different energies around the Dirac points at (K, K'). Blue cones around (K, K'): $E = \pm 0.2t_{ss}$ , green curves around (K, K'): $E = \pm 0.7t_{ss}$, red hexagon $E = \pm t_{ss}$.

(b) $p$ Orbital band structure for a tight-binding model with in-plane $p_{x,y}$ orbitals in a honeycomb lattice, in absence of $\pi$ coupling ($t_{pp\pi} = 0$) and hybridization. The energies are given in units of $t_{pp\sigma}$.

(c) Localized eigenstates for the lowest flat band in two neighboring hexagons (red and green colors, respectively). The sum of the squared amplitude of these two eigenstates is shown along the bond common to the two hexagons.

## Section G.    Calculation of the LDOS(E) spectra by using the tight-binding approximation.

Using the tight-binding model, we can obtain the eigenvectors corresponding to each energy $E_n(k)$ in the band structure, where $n$ denotes the band number. This gives the wave function distribution over the orbitals and sublattice sites for that energy. We approximate the $s$ orbitals as normalized Gaussians $Ae^{-x^2}/r$, the $p$ orbitals as $Be^{-x^2}/r \sin \varphi$ for the $p_y$ orbital and $Be^{-x^2}/r \cos \varphi$ for the $p_x$ orbital, where $A$ and $B$ are normalization constants, $r$ is proportional to the lattice size, $r = \frac{\sqrt{3}}{10} a$, with $a$ the lattice spacing, and $\varphi$ the angle with respect to the horizontal axis. Using this approximation, we can calculate the wave function $\Psi(x,y)_{E_n(k)}$. If we now wish to calculate the LDOS for an energy $E$, we can sum $\Psi(x,y)_{E_n(k)}$ over $n$ and a (dense enough) $k$ grid, where each contribution is weighted by the broadening $L[E_n(k) - E]$. Here $L(x)$ is given by $\frac{b}{\left[x^2 + \left(\frac{b}{2}\right)^2\right]}$ with $b$ the broadening of 0.04 eV.

In Fig. S8, the resulting maps and spectra are shown for a simple tight-binding model. On the lattice sites, the $s$ orbital Dirac cone is manifest, but the $p$ orbital bands have nearly zero intensity due to the nodal planes. The $p$ orbital flat bands and the $p$ orbital Dirac cone are mostly localized on the bridge sites. In this tight-binding calculation, no $s$-$p$ hybridization or $t_{pp\pi}$ are taken into account, the overlap integrals (which are small compared to the hoppings, see Table S I) have been neglected.

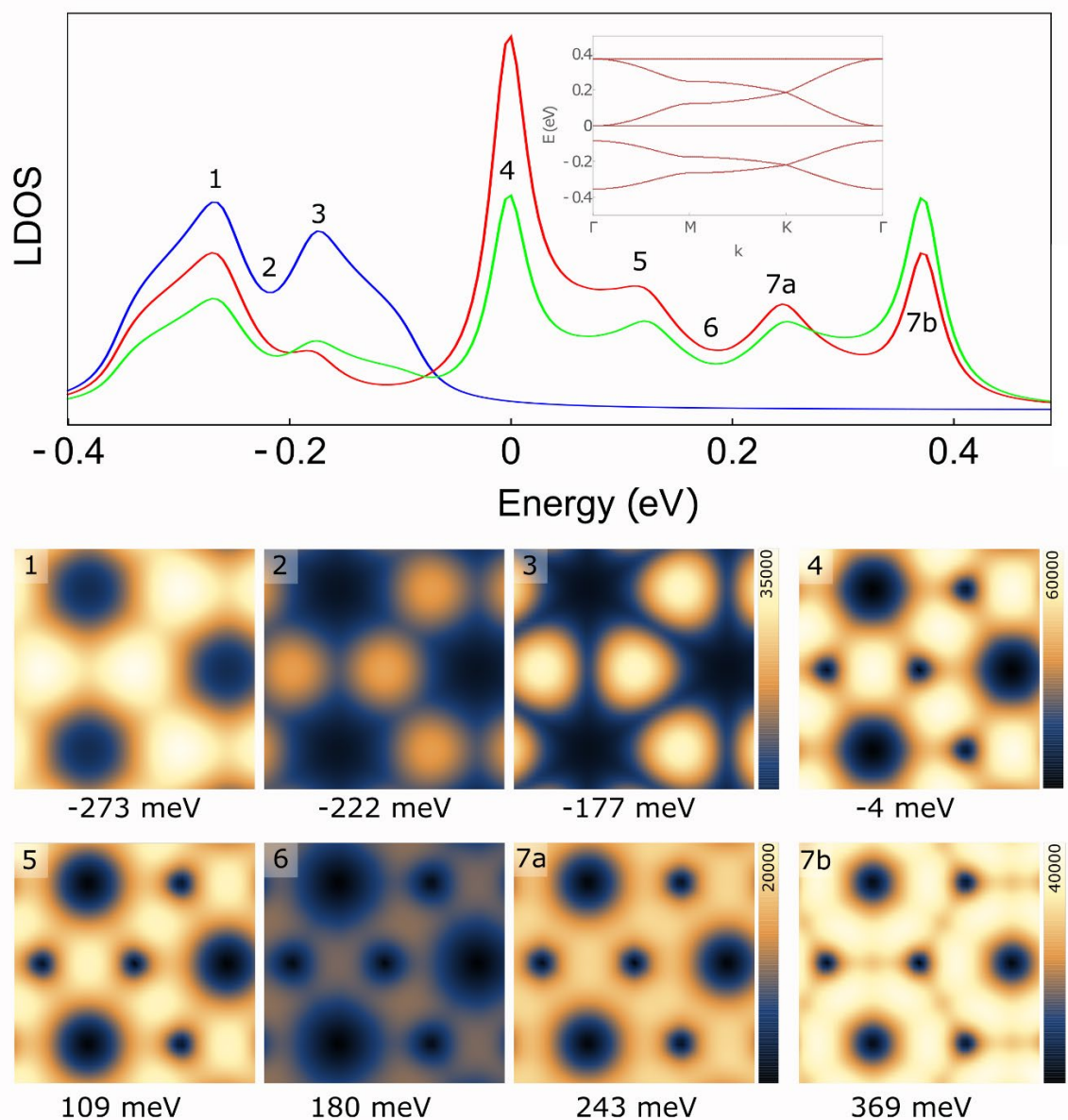

**Figure S8.** Tight-binding analysis of the spectra, band structure and wave function maps of a periodic system for the lattice with the double rosettes.

(Top) Spectra are shown for three positions in the lattice: artificial atomic sites (blue), bridge sites (red) and very close to a CO rosette (green). Inset shows the band structure.

(Bottom) LDOS maps corresponding to the interesting features in the spectra with no orbital overlap, *s-p* hybridization or $t_{pp\pi}$.

# Section H. Differential LDOS conductance maps acquired for the lattice of Fig. 2 in the energy region of the *p* orbital Dirac cone.

The spatial patterns in the energy region of the *p* orbital Dirac cone are very detailed and typical (Fig. S9(a, b)). There is high intensity close to the rosettes, very weak intensity on the sites and even weaker intensity between the sites. As a guide to the eye, a scheme is presented in Fig. S9(c) where the CO rosettes are left uncolored for clarity. At the Dirac point, the rings of high intensity around the rosettes are nearly uniform. For both peaks (M points) around the Dirac point, the high intensities form trigonal arrays around each artificial site.

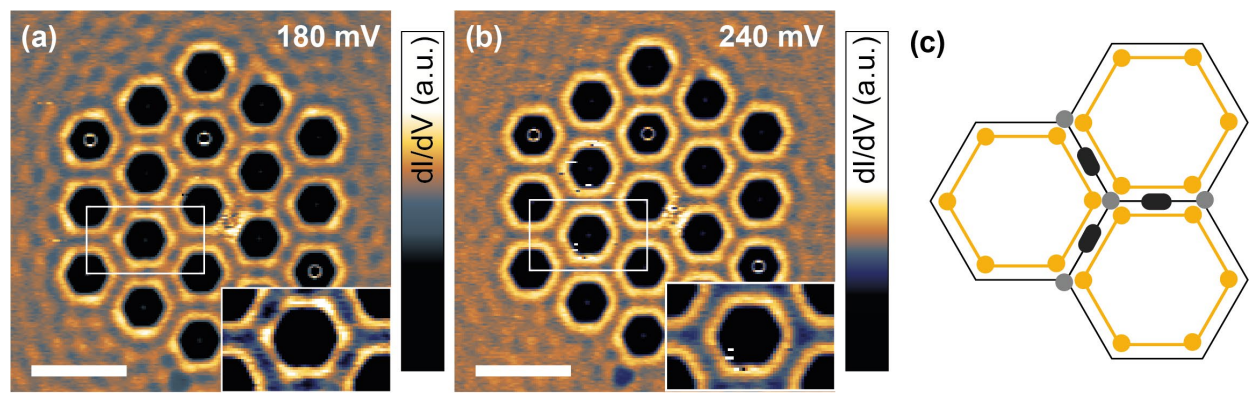

**Figure S9.** Differential LDOS conductance maps acquired for the lattice of Fig. 2 in the energy region of the *p* orbital Dirac cone.

(a) Electron probability map taken at 180 mV (lower energy maximum of *p* orbital Dirac cone). The atomic sites have a low intensity while the bridge sites have an even lower intensity. The rings of high intensity around the CO rosettes show a modulation in the intensity as well.

(b) Electron probability map taken at 240 mV (*p* orbital Dirac point). The atomic sites have a low intensity while the bridge sites have a slightly higher intensity. The insets show an enlargement. Scale bar is 5 nm.

(c) Scheme of the intensities around an artificial site for the region of *p* orbital Dirac cone. The artificial site has low intensity (light grey), the bridge sites have even lower intensity (dark grey). Each rosette is circumvented with high LDOS intensity, resulting in a triangle of high intensity around each artificial site.

**Section I. Comparison of the experimental results obtained on the artificial lattice with double-ringed rosettes (Fig. 1(c)) with the muffin-tin calculations.**

Figure S10 shows a comparison between experimental results and muffin-tin calculations. Overall, an excellent agreement is found between experimental and theoretical LDOS maps for the energy region with the $s$ orbital Dirac cone and $p$ orbital flat band, and a reasonable agreement for the region of the $p$ orbital Dirac cone.

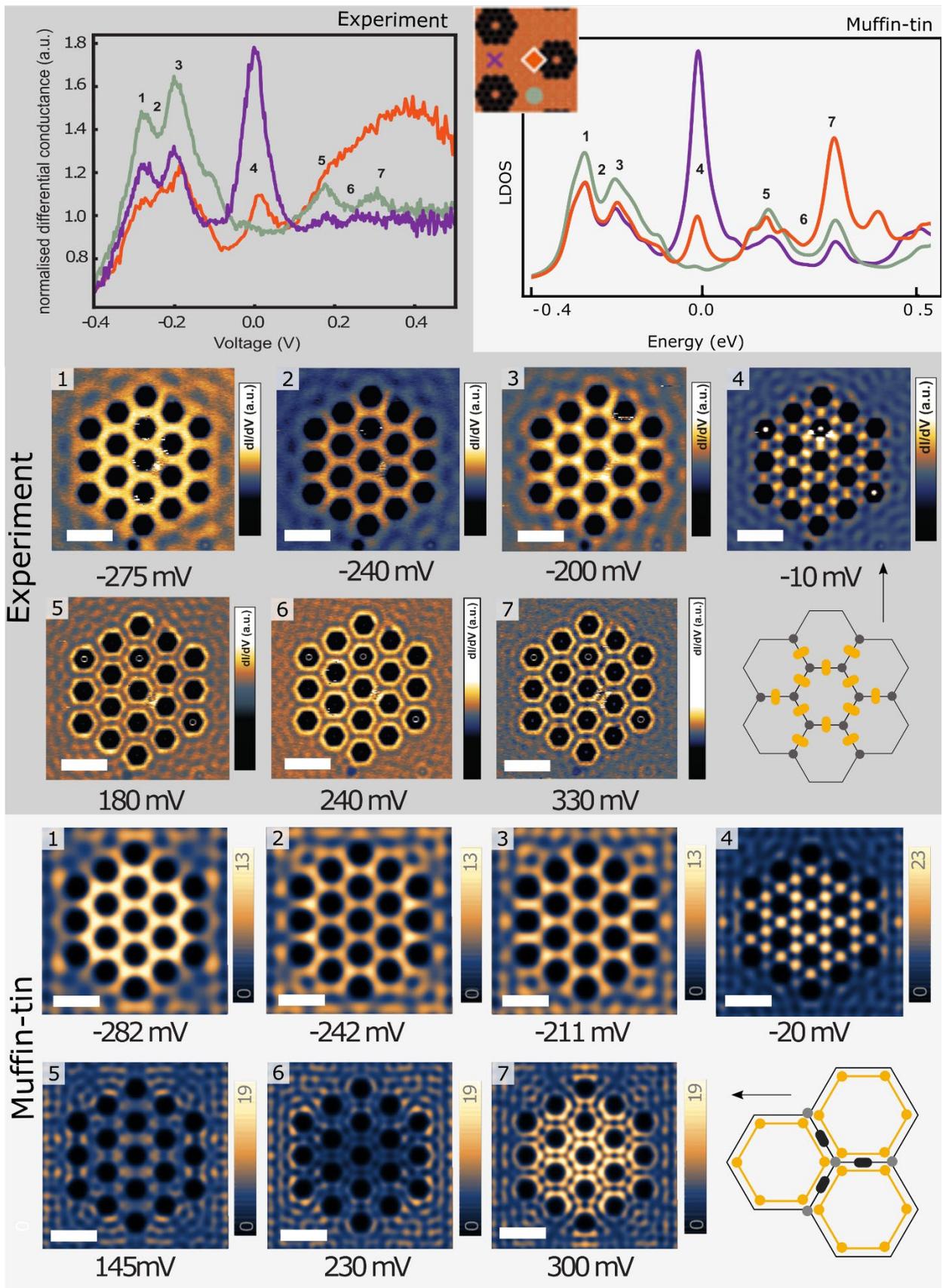

**Figure S10. Comparison of the experimental results obtained on the artificial lattice with double-ringed rosettes (Fig. 1(c)) with the muffin-tin calculations.**

First row: Experimental spectra (left) and the muffin-tin calculations (right). The numbers are related to the LDOS maps below. Notice that the orange spectrum (close to the rosettes) was taken with a different tip. Peaks 1-5 are at the same position, but have a slightly different relative intensity. This can either be due to a remaining effect of the tip change, but could, in our view, also be caused due to the fact of the very close proximity of the position of measurement to the CO rosette.  Insert: A sketch of the locations at which the spectra were taken.

Second and third row: Experimental electron probability (LDOS) maps acquired at constant-heights at the spectral features (numbers 1-7). The intriguing LDOS map at the flat band (point 4) has been summarized in a sketch that emphasizes the high electron density (yellow) between the artificial sites (grey circles on the hexagon). The high intensity regions between the sites form plaquettes around each hexagon of the honeycomb lattice.

Fourth and fifth row: LDOS maps calculated with the muffin-tin model. The maps 1-4 are in excellent agreement with the corresponding experimental results. The calculated maps 5 and 6 (first peak of the *p* orbital Dirac cone and Dirac point) show intricate and rapid oscillations in intensity that differ from those in the experimental spectrum; in addition, the calculated overall intensity is much weaker than observed experimentally.  In contrast, the calculated LDOS map 7, at the second peak of the Dirac cone, is in very reasonable agreement with the experimental result shown at 330 mV. As in our muffin-tin approach, the CO's are modelled as vertical potential barriers, the muffin-tin results in close proximity to the CO barriers become less accurate.For convenience, the detailed LDOS pattern found in the region of the *p* orbital Dirac cone is sketched again. The regions of high intensity are indicated in yellow. Each artificial atom (grey circle) is surrounded by three regions of high intensity (yellow disks) and with three regions of very low intensity (black regions at bridge sites). The regions of high intensity (yellow) form hexagonal plaquettes. The low intensity regions (bridge sites) form plaquettes around each hexagon of the honeycomb lattice. The scale bars are 5 nm.

### Section J.    A check of the uniformity of the LDOS across the artificial lattice of Fig. 2 by differential conductance spectra acquired along lines in the lattice

To check that the LDOS measurements are similar throughout the entire lattice, we obtained many spectra along a line to visualize the uniformity (Fig. S11). Periodic intensity plots show the strong reproducibility of the LDOS on different sites across the lattice. At 0 V, the high intensity on bridge sites reflecting the *p* orbital flat band is reproducibly observed. The orange sites closer to the rosettes show high intensity above 0.2 V due to the peaks of the *p* orbital Dirac cone, in line with the maps presented in the main text, Figs. 3, 4.

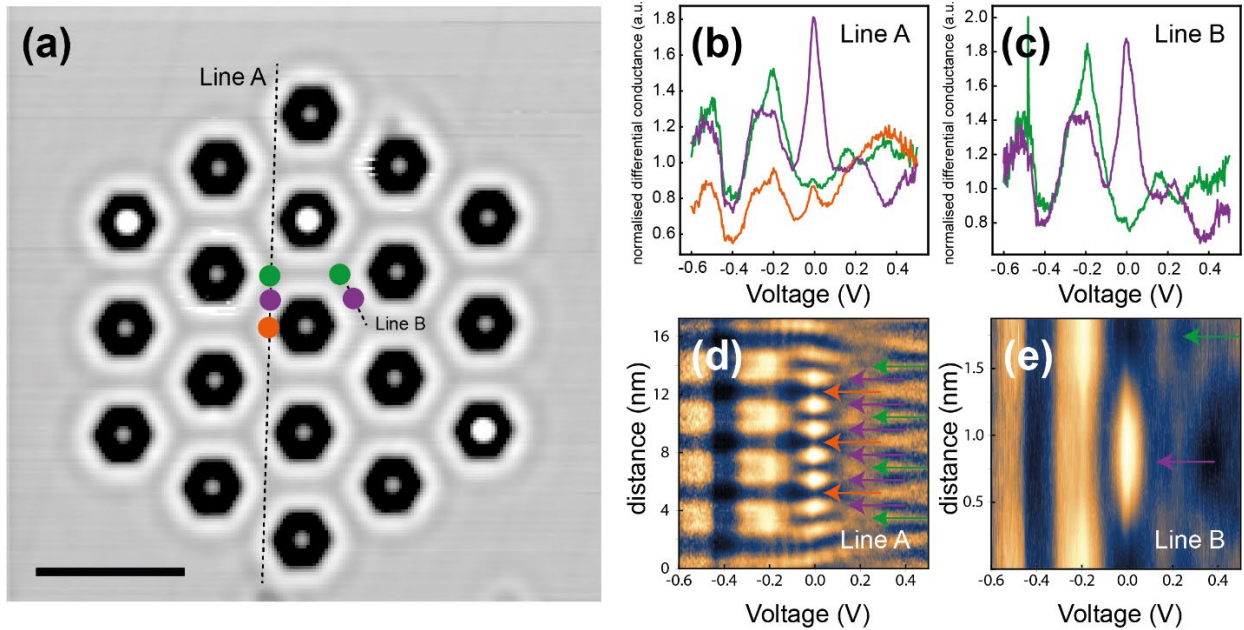

**Figure S11.** A check of the uniformity of the LDOS across the artificial lattice of Fig. 2 by differential conductance spectra acquired along lines in the lattice

(a) An STM image of the lattice with two dashed black lines indicating the line traces along which spectra were taken. 100 spectra were taken along line A, consecutively on the green lattice sites, the violet positions between the lattice sites, and orange positions closer to the CO rosettes. On line B 15 spectra were taken on green and violet sites. The scale bar is 5 nm.

(b, c) Individual representative spectra taken on lines A and B, respectively.

(d, e) Colored LDOS intensity plots obtained from all 100(15) spectra taken on line A(B), respectively, presented in a (line position – bias) frame. The arrows indicate the locations corresponding to the colored locations in panel (a).

## Section K.    Three-dimensional $E(k_x, k_y)$ diagram of the band structure of the artificial lattice in Fig. 1(c).

For the experimentally realized design presented in Fig. 1(c), we show a three-dimensional representation of the band structure in Fig. S12.

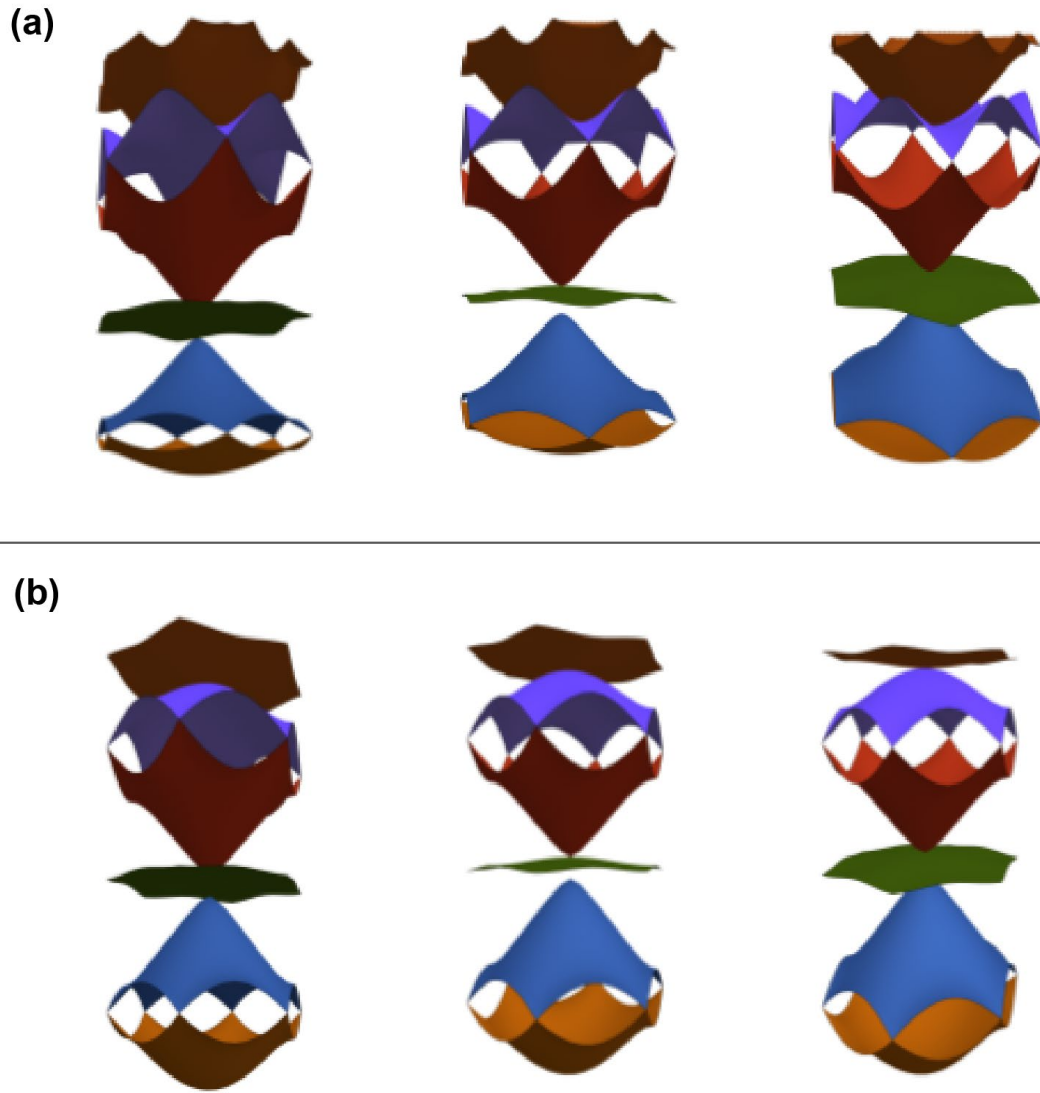

**Figure S12.** Three-dimensional $E(k_x, k_y)$ diagram of the band structure of the artificial lattice in Fig. 1(c).

(a,b) Three E(k$_x$, k$_y$) diagrams at various viewing angles of the band structure corresponding to the double-ringed rosette lattice calculated with the muffin-tin (a) and tight-binding (b) approach.